\documentclass[prb, twocolumn,superscriptaddress,nofootinbib]{revtex4-2}
\usepackage{epsfig}
\usepackage{amsmath}
\usepackage{graphicx}
\usepackage{color}
\usepackage[normalem]{ulem}
\usepackage[usenames,dvipsnames]{xcolor}
\usepackage[colorlinks=true,citecolor=Cerulean,linkcolor=RubineRed,urlcolor=Cerulean]{hyperref}
\usepackage{amsfonts,amssymb}
\usepackage[english]{babel}
\usepackage{hyperref}
\usepackage{lipsum}
\usepackage{graphics}
\usepackage{array}
\usepackage{tabu}
\usepackage{bm}
\usepackage{float}
\usepackage{setspace}

\newcommand{\be}{\begin{equation}}
\newcommand{\ee}{\end{equation}}
\newcommand{\ba}{\begin{eqnarray}}
\newcommand{\ea}{\end{eqnarray}}

\newcommand{\non}{\nonumber}
\newcommand{\bra}[1]{\langle #1|}
\newcommand{\ket}[1]{|#1\rangle}

\begin{document}

\title{Shortcuts to adiabaticity in open quantum critical systems}
\author{Shishira Mahunta}
\affiliation{Department of Physical Sciences, IISER Berhampur, Berhampur 760003, India}
\author{Victor Mukherjee}
\affiliation{Department of Physical Sciences, IISER Berhampur, Berhampur 760003, India}

\begin{abstract} 
We study shortcuts to adiabaticity (STA) through counterdiabatic driving in quantum critical systems, in the presence of dissipation. We evaluate unitary as well as non-unitary controls, such that the system density matrix follows a prescribed trajectory corresponding to the eigenstates of a time-dependent reference Hamiltonian, at any instant of time. 
The strength of the dissipator control term for the low energy states show universal scaling close to criticality. Using the example of free-Fermionic systems, and in particular the transverse-field Ising model, we show that in contrast to STA in closed quantum critical systems, here STA may require multi-body interactions terms, even away from criticality,  owing to change in entropy of the  time-dependent target state. Further, the associated heat current shows extremum, while power dissipated changes curvature, close to criticality, and analogous to unitary control, no operational cost is associated with implementation of the exact counterdiabatic Hamiltonian. We expect the counterdiabatic protocol studied here can be of fundamental importance for understanding STA in many-body open quantum systems, and can be highly relevant for varied topics involving open many-body quantum systems, such as quantum computation and many-body quantum heat engines.

\end{abstract}
\maketitle

Quantum systems driven out of equilibrium in general show non-adiabatic excitations. Consequently,  shortcuts to adiabaticity (STA) has been developed in the recent years, which provides an alternative pathway to effective adiabatic evolution, even for finite rates of driving \cite{Chen10}.  STA can aid in  quantum computation \cite{Santos15}, can be utilized for the preparation of given target states in finite times \cite{Bukov2021state, KOLODRUBETZ20171},  and also has been shown to be highly beneficial for enhancing the output of quantum thermal machines \cite{campo14more, hartmann19many}.
STAs can be engineered in both classical \cite{deffner14classical, patra17shortcuts} and quantum systems \cite{Torrontegui13} and by now have been demonstrated in a wide variety of experiments \cite{Torrontegui13,delcampo19,GueryOdelin19}.

A universal approach to implementing STA in closed quantum systems relies on counterdiabatic driving (CD) \cite{Demirplak03,Demirplak05,Demirplak08}, also known as transitionless quantum driving \cite{Berry09}. This approach is based on application of a control Hamiltonian $H_{CD}$, such that the total Hamiltonian $H_{STA} = H_0 + H_{CD}$ ensures effective adiabatic dynamics corresponding to a reference Hamiltonian $H_0$, which is modulated in time at a finite rate. 

In general, Kibble-Zurek mechanism dictates the generation of non-adiabatic excitations  in many-body quantum systems driven through quantum phase transitions \cite{zurek05dynamics, polkovnikov05universal}. However, as shown in Ref. \cite{delcampo12}, implementation of STA protocol can be particularly challenging in quantum critical systems (QCSs), owing to the presence of highly non-local control terms involving multi-spin interactions close to criticality. In addition, engineering STA through counterdiabatic driving may require detailed knowledge about the eigen spectrum of the quantum system, which can be highly non-trivial in many-body quantum systems.
These requirements can be prohibitive in many applications, such as in quantum annealing \cite{Dutta16} and quantum computing \cite{Santos15}. 
Much of the ensuing work on STA in quantum critical systems has  focused on alleviating these challenges \cite{Takahashi13,Takahashi14,delcampo15,Campbell15,mukherjee16local,Okuyama16,Bachmann17}, with an eye on applications on quantum computation and optimization \cite{Kais14,Santos15, MichaelPRXQuantum,Takahashi17,Prielinger21}.  In the recent years approximate counterdiabatic protocols have been developed, which removes the requirement of a detailed knowledge about the eigenspectrum of a system, and hence can be highly beneficial for implementation of STA in many-body quantum systems \cite{Saberi14,Sels17,Claeys19}.
The combination of digital quantum simulation of the CD control in combination with variational ansatz has led to a new class of quantum algorithms, known as digital counterdiabatic quantum algorithms \cite{Hegade21,Hegade21Portfolio,Chandarana22,Hegade22DCQO,Hegade21factorization}.

The works mentioned above focused on STA in close quantum systems involving unitary dynamics. However, developing STA in the presence of dissipation can be crucial as well, for their application potential in open quantum systems, such as for fast cooling or heating of quantum systems \cite{Alipour20,yin2022shortcuts}, or for enhancing the output of quantum heat engines \cite{revathy22bath}. 
 The development of STA in quantum open systems is in its infancy and only a few results are known to date \cite{Vacanti14,Dann19,Alipour20,Dupays20,DannTobalina20,Dupays21,Passarelli22}. A notable experiment in this context has been reported in open circuit quantum electrodynamics \cite{yin2022shortcuts}. A universal framework generalizing CD, applicable both to isolated and open quantum systems, was put forward in \cite{Alipour20}. 
This scheme 
finds applications on thermodynamics of quantum systems \cite{Alipour22}. 
This formulation demands evolution of the system density matrix $\rho(t)$ along a prescribed trajectory, for example, the instantaneous thermal state of a time-dependent reference Hamiltonian $H_0(t)$. 
The dynamics is non-unitary, and consequently implementation of the control protocol requires CD dissipator terms that govern the change of the $\rho(t)$ eigenvalues, in addition to a CD Hamiltonian. This formalism avoids the notion of adiabaticity in open systems in terms of Jordan blocks, inherent to other approaches \cite{SarandyLidar05,Vacanti14,Passarelli22}.

The application  of the counterdiabatic scheme in open quantum systems to date has been restricted to single-particle quantum systems. In view of the challenges of applying STA in closed QCS, one can anticipate non-trivial features associated with application of STA in
finite time open QCSs as well. Chartering these difficulties is the focus of this work and we utilize the paradigmatic transverse field quantum Ising model as a testbed. This is both of fundamental relevance to understanding how STAs generalize to many-body open quantum systems and for applications such as quantum annealing \cite{Dutta16,Bando20,King22}, quantum metrology \cite{Alipour14,Beau17metro,Rams18,Frerot18}, and the engineering of quantum thermodynamic devices \cite{revathy20universal,BhattacharjeeDutta20,Mukherjee21}, involving open QCSs. Notably, our analysis shows that in stark contrast to closed QCSs, STA in the presence of dissipation may involve many-body control terms even away from criticality. 

We present the STA protocol in general open quantum systems driven out of equilibrium in Sec. \ref{sec:gen}. We then focus on the transverse Ising model in the presence of a Fermionic bath Sec. \ref{secFF}; we focus on the low temperature regime in Sec. \ref{subLow}, followed by real space description of the CD control terms in Sec. \ref{subSTAreal}. We then study power and heat current associated with the STA control in Sec. \ref{secPH}. Finally, we conclude in Sec. \ref{secConcl}.

\section{General setup}
\label{sec:gen}

We consider a generic many-body system described by a time-dependent Hamiltonian $H_0(t) = \sum_n E_n(t)\ket{n_t}\bra{n_t}$, and coupled to a dissipative bath at an inverse  temperature $\beta$. Here, $E_n(t)$ is the energy of the $n$-th instantaenous energy eigenstate $\ket{n_t}$ with respect to $H_0(t)$ at time $t$, with $E_n \leq E_m$ for $n < m$; $m, n \in\mathbb{N}$. We focus on the case of a system-bath interaction such that for a $H_0(t)$ varying infinitesimally slowly with time, the system always stays in its instantaneous thermal equilibrium state with the bath, described by a Gibbs state $\rho_{G}^{(\beta)}$ (see Eq. \eqref{eq:target}), as determined by the bath parameter $\beta$. In contrast, for a $H_0(t)$ changing  at a finite rate in time, in general the system is driven away from it's instantaneous thermal equilibrium state in absence of control.  Consequently, here we evaluate the counter-diabatic control terms required to ensure that the spectral decomposition $\rho(t) = \sum_n \lambda_n(t)\ket{n_t}\bra{n_t}$ of the system density matrix can be recast in the form of the instantaneous Gibbs state $\rho_G^{(\beta)}(t)$ at all times, i.e.,
\ba
\rho(t) &=& \rho_G^{(\beta)}(t) = \sum_n \lambda_n(t)\ket{n_t}\bra{n_t};~\lambda_n(t) = \frac{e^{-\beta E_n(t)}}{Z_0(t)},\non\\
Z_0(t) &=& {\rm Tr}\left(e^{-\beta H_0(t)}\right)
\label{eq:target}
\ea
We note that the above form of Gibbs state is widely used and can be highly relevant for various topics in open quantum systems \cite{breuer02} and quantum technologies \cite{mehboudi19thermometry, hartmann19many}, and is distinct from the generalized Gibbs ensemble that may arise in closed quantum systems driven out of equilibrium \cite{vidmar16generalized}.
Following Ref. \cite{Alipour20}, we note that 
the equation of motion of $\rho(t)$ involves a generator that is not unitary, and requires both unitary and non-unitary control terms. Specifically, the trajectory $\rho(t)$ obeys the master equation
\ba
\dot{\rho}(t) &=& \mathcal{L}[\rho(t)] = -i\left[H_{\rm STA}, \rho(t)\right] \non\\
&+& \sum_{m,n}\gamma_{mn} \left(L_{mn}\rho L_{mn}^{\dagger} -\frac{1}{2}\{L_{mn}^{\dagger}L_{mn},\rho \}\right).
\label{eqmaster}
\ea
Here, $H_{\rm STA} =  H_0 + H_{CD}$, and 
\ba
H_{CD}(t) = \mathrm{i} \hbar \sum_n\left(\left|\partial_t n_t\right\rangle\left\langle n_t\left|-\left\langle n_t \mid \partial_t n_t \right\rangle\right| n_t\right\rangle\langle n_t|\right) 
\label{eq:Hcd}
\ea
is the counterdiabatic Hamiltonian with respect to $H_0(t)$  \cite{Berry09, delcampo12},  while
\ba
L_{mn}(t) = \ket{m_t}\bra{n_t};~~~~~\gamma_{mn}(t) = \frac{\dot{\lambda}_{m}(t)}{r \lambda_n(t)},
\ea
where $r$ is the rank of $\rho(t)$.


One can evaluate the coefficient $\gamma_{mn}$ corresponding to a transition from the $n$-th energy state to the $m$-th energy state ($m, n = 1, 2, \hdots$) as (see Appendix \ref{appA} for details)
\ba 
\gamma_{mn} = \frac{\beta e^{-\beta \left(E_m - E_n\right)} \sum_l e^{-\beta E_l} \left(\dot{E}_l - \dot{E}_m \right)}{r\sum_l e^{-\beta E_l}}. 
\label{eqgammn}
\ea
As expected, the low-temperature limit ($\beta \gg \left(E_m - E_n\right)^{-1}$ $\forall~m, n$) is associated with high rates of transitions to the lower energy states, as signified by $|\gamma_{mn}| \gg  |\gamma_{nm}|$ for $E_n > E_m$ (see Eq. \eqref{eqgammn}).
For the lower energy states $m = 1,~n=2$ in the low temperature limit we get
\ba
\gamma_{12} &\approx&  \frac{\beta  \left(\dot{E}_2 - \dot{E}_1 \right)}{r\left[1 + e^{-\beta \left(E_2 - E_1\right)}\right]}, 
\ea
where we have considered $\exp\left[-\beta E_l\right] \to 0$ for $l \geq 3$. 
Further, we have $\left(E_2 - E_1 \right) \sim \mu^{\nu z}$ close to criticality, where $\mu ~(\gg L^{-1/\nu}$) denotes the distance from the quantum critical point; here $\nu$ ($z$) is the corresponding correlation length (time) exponent, while $L$ denotes the size of the system \cite{sachdev99quantum}. Consequently in this regime one has
\ba
\gamma_{12} \sim \frac{\beta \nu z \mu^{\nu z -1} \dot{\mu}}{r\left[1 + \exp\left(-\beta \mu^{\nu z}\right)\right]},
\label{gammscaling}
\ea 
signifying a diverging $\gamma_{12}$ close to criticality for $\nu z < 1$, while $\nu z > 1$ results in $\gamma_{12} \to 0$. We note that the vanishing $\gamma_{12}$ close to criticality for $\nu z > 1$ is in stark contrast to unitary dynamics, where in general the strength of the counter-diabatic Hamiltonian diverges close to criticality. This can be related to the fact that for unitary dynamics,  the counter-diabatic Hamiltonian is associated with a sudden change of the ground state of the Hamiltonian close to criticality. In contrast, here $\gamma_{12}$ is related to $\left(\dot{E}_2 - \dot{E}_1 \right)$, which can be small for $\mu \to 0$ and $\nu z > 1$.

We note that in spite of the wide applicability of the above form of the Gibbs steady state \eqref{eq:target}, cases may arise where depending on the details of setup, there might be additional constraints determining the steady state of a system. For example,  the steady state of a free-Fermionic system  coupled to a Fermionic bath may  be describable by $k$-dependent Gibbs states instead \cite{keck17dissipation, revathy20universal, revathy22bath}. In the next section we focus on such a scenario, and study STA in a free-Fermionic system in the presence of dissipation.

\section{STA in free-Fermionic systems}
\label{secFF}

We consider a family of $d$-dimensional free-Fermionic systems described by the Hamiltonian
\ba \label{Eq: Free_Ham}
 H_0(t) = \sum_k \Vec{\Psi}^\dagger_k \tilde{H}_k(\zeta(t)) \Vec{\Psi}_k.
\ea
Here $\tilde{H}_k(\zeta(t)) =  \Vec{b}_k( \zeta(t)) .  \Vec{\sigma}_k$,  $\zeta(t)$ is a scalar parameter which is being changed in time in order to drive the system across a quantum critical point, and $\Vec{\Psi}_k = ( c^\dagger_{1k} , c^\dagger_{2k})$ and $\Vec{\sigma}_k = ( \sigma^x_k , \sigma^y_k, \sigma^z_k )$  represent respectively the Fermionic operators and the Pauli matrices acting on the $k$-th momentum mode. The function $ \Vec{b_k(\zeta(t))} = (b^x_k , b^y_k , b^z_k )$ is system specific, and depending on its explicit form, $H_0(t) $  can describe various systems including XY and Ising model in $d = 1$  \cite{bunder99effect,lieb61two, pfeuty70the, sachdev99quantum}, and Kitaev model in $d = 1$ and $d = 2$ dimensions \cite{kitaev06anyons, KrishnenduPhysRevLett2008, mondal08quench, DungPhysRevLett2007}.
The  energy gap $\Delta_k  = 2 \sqrt{\left(b^x_k\right)^2 + \left(b^y_k\right)^2  + \left(b^z_k\right)^2}$ between the ground state and the first excited state vanishes at the quantum critical point $\zeta = \zeta_c$, for the critical mode $k = k_c$. We consider a free-Fermionic bath, which, for a $\zeta(t)$ varying infinitesimally slowly in time, takes the $k$ mode to its instantaneous Gibbs state 
\ba
\rho_{G,k}^{(\beta)} &=& \sum_{n_k} \lambda_{n,k}(t) \ket{n_{k,t}} \bra{n_{k,t}};\non\\
\lambda_{n,k} &=& \frac{\exp\left[-\beta E_{n,k}(t)\right]}{Z_k(t)}.
\label{eqGibbsk}
\ea
Here $\ket{n_{k,t}}$  ($E_{n,k}(t)$) denotes the $n$-th instantaneous eigenstate (eigenenergy) for the mode $k$, and $Z_k(t)$ is the corresponding partition function,  at time $t$. For example, this kind of $k$-dependent Gibbs state can arise in the presence of Fermionic baths, and can be relevant for quantum computation \cite{keck17dissipation}, studying dynamical quantum phase transitions in the presence of dissipation \cite{bandyopadhyay18exploring}, and for designing quantum engines \cite{revathy20universal, revathy22bath}. Such Fermionic baths are non-local in real space, while being local in the $k$ space, thereby giving rise to the Gibbs state Eq. \eqref{eqGibbsk} \cite{keck17dissipation, revathy20universal, revathy22bath}.

Next we focus on the specific example of a particular free-Fermionic system, namely the transverse Ising model, given by the Hamiltonian
\ba
H_0(t) = -\frac{1}{2}\left[\sum_j \sigma_j^x \sigma_{j+1}^x + h(t)\sum_j\sigma_j^z\right].
\ea
This model exhibits a quantum phase transition between a paramagnetic phase $|h|>1$ and a ferromagnetic phase ($|h|<1$) at the critical points $h_c = \pm 1$ \cite{bunder99effect,lieb61two, pfeuty70the, sachdev99quantum}.
One can make use of the Jordan-Wigner transformation followed by a Fourier transformation to arrive at the free-Fermionic form
\ba
\tilde{H}_k(t) &=& \begin{bmatrix}
 (t/\tau + \cos k) & 0 & 0 & i\sin k \\ 0 & 0 & 0 & 0\\ 0 & 0 & 0 & 0\\ -i\sin k & 0 & 0 & -(t/\tau + \cos k)
 \label{Hkmatrix}
\end{bmatrix},
\ea
in the basis $\{\ket{0_k,0_{-k}}, ~\ket{0_k,1_{-k}} = c_{-k}^{\dagger}\ket{0_k,0_{-k}}, ~\ket{1_{k},0_{-k}} = c_{k}^{\dagger}\ket{0_k,0_{-k}}, ~\ket{1_k,1_{-k}} = c_{k}^{\dagger}c_{-k}^{\dagger}\ket{0_k,0_{-k}}$\}, 
where we have assumed $h(t) = t/\tau$ \cite{lieb61two}. Here we note that the presence of nonunitary dynamics can allow the transition between singly excited states such as $\{\ket{0_k,1_{-k}}, \ket{1_{k},0_{-k}} \} $\cite{keck17dissipation}. Consequently, we represent $\tilde{H}_k(t)$ by the full  $4 \times 4 $  Hamiltonian matrix \eqref{Hkmatrix}.  We assume the $k$th mode starts in its instantaneous thermal equilibrium (Gibbs) state at $t \ll \tau$. However, as the transverse field $h(t)$ is changed, one can expect the system to go out of thermal equilibrium for finite rates of $\tau$. Consequently, here we evaluate the Lindblad control protocol which ensures that the system remains in its  instantaneous local Gibbs state $\rho_k = \rho_{G,k}^{(\beta)}$, given by
\ba
\rho_{G,k}^{(\beta)} &=& \begin{bmatrix}
\frac{e^{-\beta \epsilon_k}}{Z_k}& 0 & 0 & 0 \\ 0 & \frac{1}{Z_k} & 0 & 0\\ 0 & 0 & \frac{1}{Z_k} & 0\\ 0 & 0 & 0 & \frac{e^{\beta \epsilon_k}}{Z_k}
\end{bmatrix},
\label{rhoth}
\ea
in the energy eigenbasis for the k-th mode, at all times. In this expression,  
 $\epsilon_k = \sqrt{\left(t/\tau + \cos k \right)^2 + \sin^2 k}$ and $Z_k = 2 + e^{-\beta \epsilon_k} + e^{\beta \epsilon_k}$. The energy gap vanishes at the critical point $h(t) = \pm 1$ for $k = 0, \pi$ \cite{lieb61two, dutta15quantum}. Furthermore, free-Fermionic nature of the system implies $\rho(t) = \bigotimes_k \rho_k(t)$, thus allowing us to treat each $k$ mode independently.

Following Sec. \ref{sec:gen}, we have 
\ba
L_{mn}^{(k)}(t) = \ket{m_t^{(k)}}\bra{n_t^{(k)}},
\label{lmn}
\ea
where $\ket{m^{(k)}_t}$ denotes any of the four instantaneous eigenstates for the $k$-th mode, given by
\ba
\ket{1_t^{(k)}} &=& \frac{\alpha^{(k)}_t \ket{1_k, 1_{-k}} + \ket{0_k, 0_{-k}}}{\sqrt{|\alpha^{(k)}_t|^2 + 1}}, \non\\
\ket{2_t^{(k)}} &=& \ket{1_k, 0_{-k}},\non\\
\ket{3_t^{(k)}} &=&  \ket{0_k, 1_{-k}},\non\\
\ket{4_t^{(k)}} &=& \frac{\theta^{(k)}_t \ket{1_k, 1_{-k}} + \ket{0_k, 0_{-k}}}{\sqrt{|\theta^{(k)}_t|^2 + 1}}, \non\\
\alpha^{(k)}_t &=& i\frac{\left(t/\tau + \cos k - \epsilon_k\right)}{\sin k},  \non\\
\theta^{(k)}_t &=& i\frac{\left(t/\tau + \cos k + \epsilon_k\right)}{\sin k}. 
\label{levelsk}
\ea
Here, $\ket{1_t^{(k)}}$ is the ground state, $\ket{2_t^{(k)}}$ and $\ket{3_t^{(k)}}$ are degenerate second and third excited states, and $\ket{4_t^{(k)}}$ is the highest excited state. The explicit forms of the Lindblad operators and the corresponding coupling rates are derived given in Appendix \ref{appB}.
We note that $\gamma_{mn}^{(k)}$ denotes the rate of transition from the $\ket{n^{(k)}}\bra{n^{(k)}}$ state to the state $\ket{m^{(k)}}\bra{m^{(k)}}$, for the mode $k$. \\

\subsection{Low temperature regime}
\label{subLow}

In this section we consider the low-temperature limit of $\beta \epsilon_k \gg 1$ $\forall~k$. In this regime the instantaneous Gibbs state is close to the instantaneous ground state $\ket{1_t^{(k)}}$ for all times. Consequently all the $\gamma_{mn}^{(k)}$'s denoting excitation to higher energy states are exponentially small ($m > n$, see Eq. \eqref{eq:gamk}), while the non-zero rates signifying transitions to the lower or degenerate energy states are given by
\ba
\gamma_{12}^{(k)} &=& \gamma_{13}^{(k)} \approx \frac{\beta \left(t/\tau + \cos k \right)}{2\tau\epsilon_k}\non\\
\gamma_{23}^{(k)} &=& \gamma_{32}^{(k)} = -\frac{\beta \left(t/\tau + \cos k \right)}{4\tau \epsilon_k} \non\\
\gamma_{34}^{(k)} &=& \gamma_{24}^{(k)}  = -\frac{\beta e^{\beta \epsilon_k}\left(t/\tau + \cos k \right)}{4\tau \epsilon_k}\non\\
\gamma_{14}^{(k)} &=& \frac{\beta e^{\beta \epsilon_k} \left(t/\tau + \cos k \right)}{2 \tau\epsilon_k}
\label{eq:gamk2}
\ea
\begin{figure*}
    \centering
    \includegraphics[width= \textwidth]{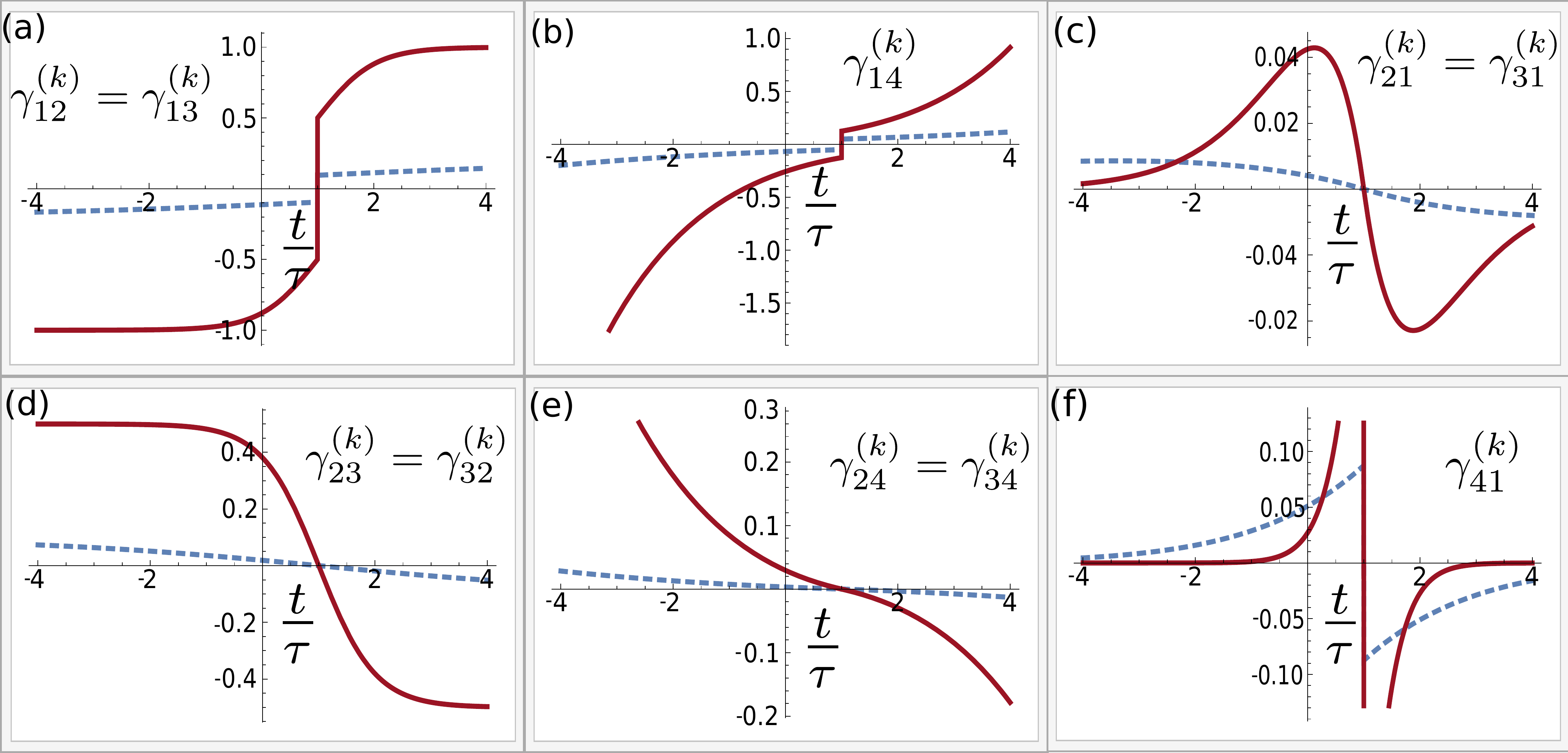}
    \caption{Plot showing  the variation of $\gamma_{mn}^{(k)}$ with time for $ k = \pi $ (See Eq. \ref{eq:gamk}) . The red solid line refers to the low temperature regime, $\beta = 1.7$ and the blue dashed line refer to the high temperature regime, $\beta = 0.25$. As shown here, $\gamma_{mn}^{(k)}(t)$'s change sign and may show non-analytic behavior close to the quantum critical point at $ t/\tau = 1$.}
    \label{fig:Close_Criticality}
\end{figure*}

We show the variations of $\gamma_{mn}$'s with time for the critical mode $k = \pi$ in Fig. \ref{fig:Close_Criticality}. As shown in Fig. \ref{fig:Close_Criticality}, the $\gamma_{mn}$'s change sign, and may show show non-analytic behavior at  $t/\tau = -\cos \pi = 1$. The time dependence of the $\gamma_{mn}$'s follow from the temporal variation of the energy eigenvalues (see Appendix \ref{appC}). A positive (negative) $\gamma_{mn}$ signifies transfer of population from the $n$-th ($m$-th) energy state to the $m$-th ($n$-th) energy state; this in turn results in the $\gamma_{mn}$'s changing sign at $t/\tau = -\cos k$ (see Eq. \eqref{eq:gamk2}), thus ensuring that the $k$ th mode stays in it's instantaneous Gibbs state at all times.   Further, $\gamma_{m4}^{(k)}(t)$ diverges for large $|t|/\tau$, implying high rates of population transfer from the highest excited state $\ket{4_t^{k}}$ to other states for large $|h(t)|$, as expected.

\subsection{STA in real space}
\label{subSTAreal}

\begin{figure}[h]
         \centering
         \includegraphics[width = \linewidth]{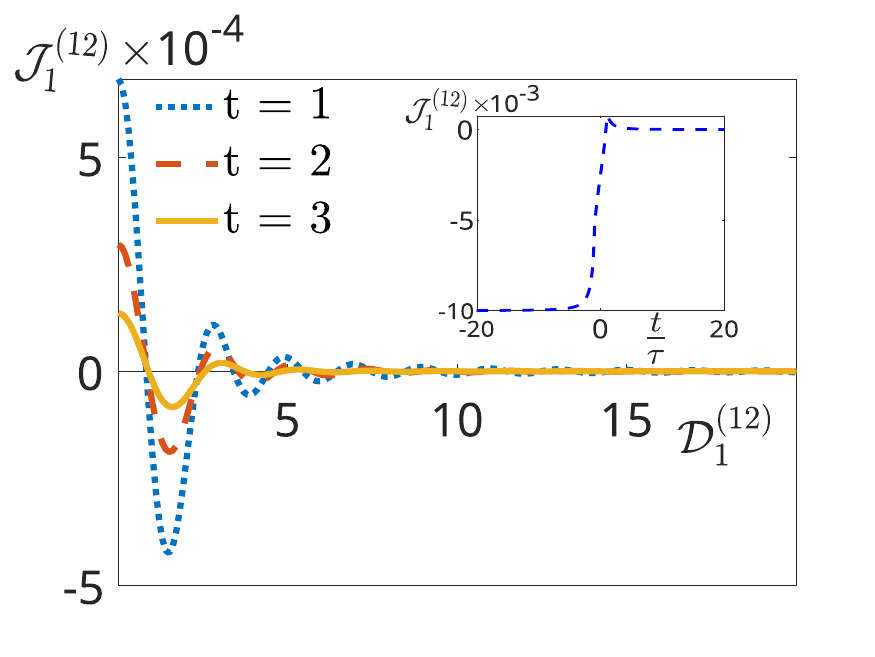}
         \caption{ The main plot shows variation of  $\mathcal{J}^{(12)}_1$ as a function of $\mathcal{D}_1^{(12)} = \left(n + m + q \right) - \left(p+r+s\right)$ and the inset shows variation of $\mathcal{J}^{(12)}_1$ as a function of $\frac{t}{\tau}$ when $\mathcal{D}_1^{(12)}= 0$. Here $\tau = 1$, $L = 1000$, $\beta = 0.01$ (see Eqs. \eqref{eqT1} and \eqref{eq:J34}).} 
        \label{fig:JD}
\end{figure}

As discussed above, the Lindblad operators are relatively simple in the momentum space, where they involve only $\pm k$ modes (see Eqs. \eqref{lmn} and \eqref{levelsk}). In order to arrive at the spin space representation of the control terms, without loss of generality we focus on $\sum_k\gamma_{12}^{(k)}(L_{12}^{(k)})^{\dagger}L_{12}^{(k)} \rho_k$, which appears in the master equation Eq. \eqref{eqmaster} for a single $k$ mode, as a representative term. This will include terms of the form
\ba
T_1^{(12)} = \sum_k A^{(12)}_{1k} c_k^{\dagger} c_{-k} c_k c_k^{\dagger} c_{-k}^{\dagger} c_k \rho_k, 
\ea
where 
\ba
A^{(12)}_{1k} = \frac{\gamma_{12}^{(k)} |\alpha^{(k)}_t|^2}{|\alpha_t^{(k)}|^2 + 1}.
\label{eq:A1k}
\ea
One can show that in the spin space, this will result in multi-body interaction terms of the form (see Appendix \ref{appreal} for details)
\begin{widetext}
\ba
\chi_1  &=&   \frac{\left(-1 \right)^{\mathcal{S} -1}}{L^3} \sum_{\{n,m,p,q,r,s\}} \left(\Pi_{j = p}^{n-1} \sigma_j^z \right)   \left(\Pi_{j = m}^{r-1} \sigma_j^z \right) \left(\Pi_{j = s}^{q-1} \sigma_j^z \right)  \sigma_n^{+}  \sigma_p^{-}  
\sigma_r^{+}  \sigma_m^{-}  \sigma_q^{+} \sigma_s^{-} \mathcal{J}^{(12)}(n,m,p,q,r,s) \rho_k,\non\\
\mathcal{J}_1^{(12)}(n,m,p,q,r,s) &\equiv& \mathcal{J}_1^{(12)}(\mathcal{D}_1^{(12)}) = \sum_k A_{1k}^{(12)} e^{i\mathcal{D}_1^{(12)} k},
\label{eqT1}
\ea
\end{widetext}
where we have taken into account $A_{1k}^{(12)} = A_{1(-k)}^{(12)}$, and assumed $\{n,m,p,q,r,s\}$ such that  $n > p > r > m > q > s$. Here $\mathcal{S} =  n + m + q + p+r+s$ and $\mathcal{D}_{1}^{(12)} = \left(n + m + q \right) - \left(p+r+s\right)$. Further, in the limit of large  $L$ Eq. \eqref{eqT1} reduces to 
\ba
\mathcal{J}_1^{(12)}(\mathcal{D}_1^{(12)}) \approx  \frac{2}{\pi} \int_0^{\pi}A_{1k}^{(12)} \cos\left(\mathcal{D}_1^{(12)} k\right) dk.
\label{eq:J12}
\ea
The terms on the r.h.s. of Eq. \eqref{eq:J12} can be expected to vanish in the limit of large $\mathcal{D}_{1}^{(12)}$. This is also verified by  Fig. \ref{fig:JD}, where it is shown that $\mathcal{J}_1^{(12)}$ decreases rapidly with $\mathcal{D}_1^{(12)}$, implying one can restrict $\mathcal{D}_1^{(12)}$ to small values only. However,  we get $x = (n - p), (r - m), (q-s)$ particle interactions in $\chi_1$ for a finite $\mathcal{D}_1^{(12)}$, where $0 \leq x \leq N$.  Moreover, these multi-particle interactions persist even far away from criticality, as well as  for $\mathcal{D}_1^{(12)} = (n - p) + (q-s) - (r - m) = 0$, as signified by non-zero values of $\mathcal{J}_1^{(12)}(\mathcal{D}_1^{(12)} = 0)$ for $t/\tau \ll -1$ in Fig. \ref{fig:JD}, or non-zero values of $\mathcal{J}_1^{(34)}(\mathcal{D}_1^{(34)} = 0)$, corresponding to the $L_{34}^{(k)}$ term(see Appendix \ref{appreal}), for $t/\tau \gg 1$ in Fig. \ref{fig:JD134}. This can be a signature of the fact that in contrast to unitary evolution, STA in open quantum systems requires control of entropy even away from criticality, which in turn necessitates multi-particle interactions in the control terms.

\section{Power and heat current under STA protocol}
\label{secPH}

In this section, we focus on the costs of the STA protocol, as determined by the corresponding heat current $\mathcal{I}_{STA}$ and power $\mathcal{P}_{STA}$. 

\subsection{Heat current}
The total heat current for a generic system in the presence of STA is given by \cite{alicki79The}
\ba
\mathcal{I}_{STA} &=& Tr( \dot{\rho} H_{STA}) \non\\ 
&=& Tr\left(-i[\rho, H_{STA}]H_{STA} + \mathcal{L}(\rho)H_{STA}\right).
\ea
The cyclic property of trace implies ${\rm Tr}\left([\rho, H_{STA}]H_{STA}\right) = 0$, thus ensuring that  $\mathcal{I}_{STA}$ can be written as a sum of bare and counterdiabatic parts (see Appendix \ref{secheat}):
\ba 
\mathcal{I}_{STA} &\equiv& \mathcal{I}_{0} + \mathcal{I}_{CD} \non\\
\mathcal{I}_{0} &=& {\rm Tr}\left[\mathcal{L}(\rho)H_{0}\right];\non\\
\mathcal{I}_{CD} &=&  {\rm Tr}\left[\mathcal{L}(\rho)H_{CD}\right].
\label{eq:heat1}
\ea
Following  Eq. \eqref{eq:target}, we get 
\ba
\mathcal{I}_0 &=& {\rm Tr} \left ( \mathcal{L}(\rho) H_0 \right ) \non \\
&=&  \text{Tr}\left ( \left ( \sum_m \dot{\lambda}_m \ket{m_t} \bra{m_t} \right) \left(\sum_n E_n \ket{n_t} \bra{n_t} \right)\right) \non \\
&=&  \text{Tr} \left( \sum_{m,n} \dot{\lambda}_m E_n \ket{m_t}\bra{n_t} \delta_{m,n}\right) \non \\
&=& \sum_m \dot{\lambda}_m E_m. 
\label{eq:J0}
\ea 
Where we have used $\mathcal{L}(\rho) = \sum_m \dot{\lambda}_m \ket{m_t}\bra{m_t}$\cite{Alipour20}. 
On the other hand, for the counterdiabatic term we have (see Eq. \eqref{eq:Hcd}): 
\ba
\mathcal{I}_{CD} &=& \text{Tr}(\mathcal{L}(\rho) H_{CD} ) \non\\
    &=&  \text{Tr}\Bigg( \left(\sum_m \dot{\lambda}_m \left | m_t \rangle  \right  \langle m_t  |\right ) \Bigg( \mathrm{i} \hbar \sum_n(|\partial_t n_t\rangle\langle n_t| \non \\
     &-& \langle n_t \mid \partial_t n_t \rangle n_t\rangle\langle n_t| ) \Bigg) \Bigg)= 0
    \label{eq:heat2}
\ea
thus showing that the $H_{CD}$ is not associated with any heat current.(see Eq. (\ref{eq:Icd}) for a detailed derivation)

\begin{figure}
    \centering
    \includegraphics[width=0.5\textwidth]{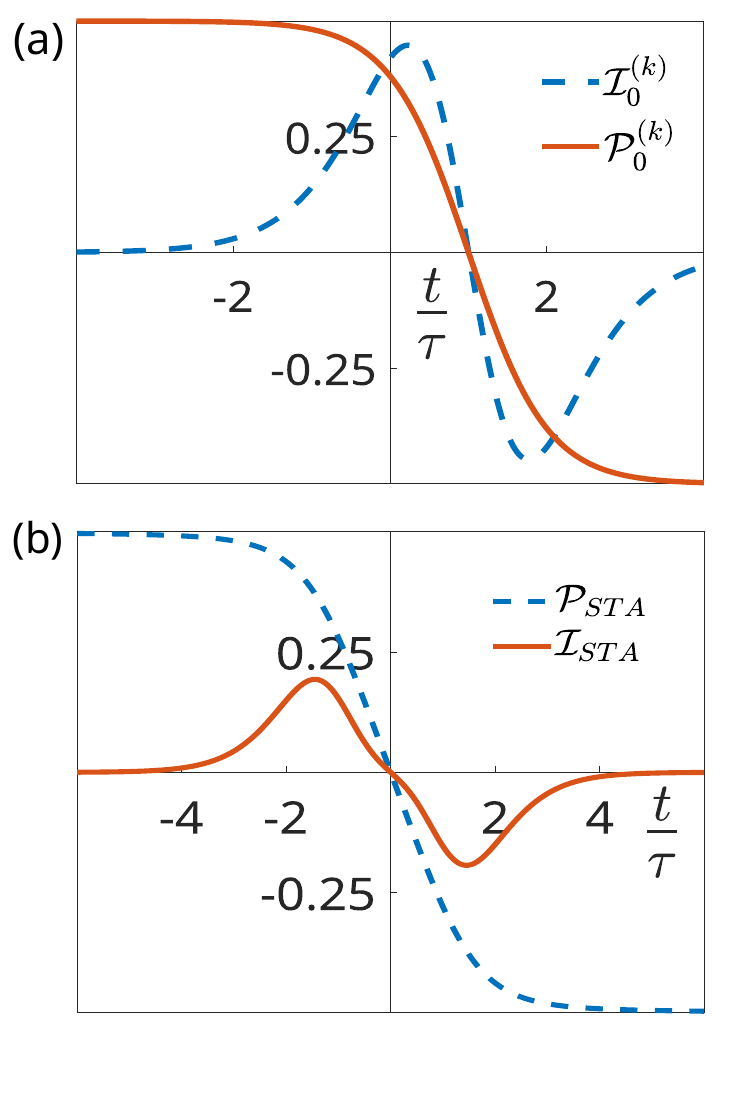}
    \caption{(a) Variation of heat current (Cf. Eq. \eqref{eqIk}) and power (Cf. Eq, \eqref{eqP0k}) with time for the transverse Ising model, for $k = \pi $. (b) Variation of  total current $ \mathcal{I}_{STA}$ and total power $\mathcal{P}_{STA}$ with time for the transverse Ising model in the thermodynamic limit. Here $\tau = 1, \beta = 2$.}
    \label{fig:Total_power_current}
\end{figure}

In the case of the transverse Ising model considered here, one has $\mathcal{I}_0 = \frac{1}{\pi}\int_0^{\pi} \mathcal{I}_{0}^{(k)} dk$, where (see Sec. \ref{secFF} and Eq. \eqref{eq:J0}), 
\ba 
\mathcal{I}_0^{(k)} &=& \sum_m \dot{\lambda}^{(k)}_m E^{(k)}_m = - \dot{\lambda}^{(k)}_1 (t) \epsilon_k + \dot{\lambda}^{(k)}_4(t) \epsilon_k \non\\ 
&=& -\frac{ \beta \left ( t/\tau + \cos(k) \right )  }{ \tau \left ( 1 + \cosh( \beta \epsilon_k \right )}.
\label{eqIk}
\ea
As shown in Eq. \eqref{eqIk} and Fig. \ref{fig:Total_power_current}(a), $\mathcal{I}_0^{k}$  is positive when the energy levels approach each other for $t/\tau < -\cos k$, signifying heat flow to the system. In contrast, the heat flow direction is reversed ($\mathcal{I}_0^{k} < 0$) for $t/\tau > -\cos k$, when the energy gaps diverge from each other. Consequently, the total heat current $\mathcal{I}_{STA} = \mathcal{I}_0$ is positive (negative) for $t < 0$ ($t > 0$), shows extremum values  close to criticality $t/\tau = \pm 1$, and approaches zero for  $|t/\tau| \gg 1$, when  the steady state does not change appreciably with time (see Fig. \ref{fig:Total_power_current} (b)).

\subsection{Power}
\label{secpow}

We now focus on the power associated with the control. As before, we
express the total power $\mathcal{P}_{STA}$  by a sum of two components  \cite{alicki79The}:
\ba
\mathcal{P}_{STA} &=& \text{Tr}\left[ \rho(t) \dot{H }_{STA} \right] = \mathcal{P}_{0} + \mathcal{P}_{CD};\non\\
\mathcal{P}_{0} &=&\text{Tr}\left[ \rho(t) \dot{H }_0 \right]  ~ \text{and}~ \mathcal{P}_{CD} = \text{Tr}\left[ \rho(t) \dot{H }_{CD} \right]
\ea

One can evaluate $\mathcal{P}_0$ as follows:
\ba
   \mathcal{P}_0 &=& \text{Tr}\left[ \rho \dot{H}_0(t) \right ]\non\\
    &=& \text{Tr} \left[  \sum_n \lambda_n | n_t \rangle \langle n_t | \left ( \frac{d}{dt}\sum_n E_n | n_t \rangle \langle n_t | \right)  \right ] \non \\ 
    &=& \text{Tr} \left[ \sum_n \lambda_n | n_t\rangle \langle n_t| \sum_n \dot{E}_n  | n_t \rangle \langle n_t |  \right] \non \\
    &+& \text{Tr} \Big[ \sum_n \lambda_n | n_t\rangle \langle n_t| E_n ( \langle n_t \mid \dot{n}_t \rangle \non\\&+& \langle \dot{n}_t \mid n_t \rangle )\Big]
    \label{eqP0}
\ea
The normalisation condition $\langle n_t | n_t \rangle = 1 $ implies $\langle n_t | \dot{n}_t \rangle = - \langle \dot{n}_t| n_t \rangle$. Therefore, $\mathcal{P}_0$  for a generic system reads 
     \ba 
     \mathcal{P}_0 &=& \text{Tr} \left[  \sum_n \lambda_n | n_t \rangle \langle n_t | \sum_n \dot{E}_n  | n_t \rangle \langle n_t |   \right ] \non\\
     &=& \sum_n \lambda_n \dot{E}_n.
     \ea 

On the other hand, as we discuss in detail in Appendix \ref{secheat}, it can be shown that $\mathcal{P}_{CD} = 0$. This is reminiscent of unitary dynamics, where it has been shown that work associated with exact counterdiabatic Hamiltonian is zero \cite{hartmann20multispin}.  

In case of the transverse Ising model we have  $\mathcal{P}_{STA} = \mathcal{P}_0 =  \frac{1}{\pi}\int_0^{\pi} \mathcal{P}_0^{(k)} dk$, where
    \ba 
    \mathcal{P}_0^{(k)} &=& \sum_n \lambda^{(k)}_n \dot{E}^{(k)}_n = -\lambda_1 \dot{E}_1^{(k)} + \lambda_4 \dot{E}_4^{(k)} \non\\ 
    &=& \frac{t/\tau + \cos(k) }{ 2 \tau \epsilon_k Z_k(t) } \left [ e^{ -\beta \epsilon_k}- e^{  \beta \epsilon_k}\right ] \non \\
    &=& - \frac{ (t/\tau + \cos(k) ) \tanh( \beta \epsilon_k /2  ) }{2 \tau  \epsilon_k }.
    \label{eqP0k}
\ea

As shown in Fig. \ref{fig:Total_power_current}(b), $\mathcal{P}_{STA}$ saturates to constant values far away from criticality, owing to the fact that the steady state does not change appreciably with time for $|t|/\tau \gg 1$ (see Eqs. \eqref{rhoth} and \eqref{levelsk}).

As discussed above, the heat current $\mathcal{I}_{CD}$ as well as the power $\mathcal{P}_{CD}$ vanish, signifying the absence of any operational cost associated with the counterdiabatic Hamiltonian $H_{CD}$. However, we note that the heat current $\mathcal{I}_{0}$ and power $\mathcal{P}_0$ dissipated are still non-zero for the bare Hamiltonian (see Fig. \ref{fig:Total_power_current}(b)). Furthermore, additional implementational costs may arise in a specific experimental setups used for realizing the STA protocol \cite{hartmann20multispin}.
\section{Conclusions}
\label{secConcl}

We have studied shortcuts to adiabaticity in quantum critical systems driven out of equilibrium in the presence of a dissipative environment. We have derived non-Markovian Lindblad operator control terms required to ensure that the system always stays in its instantaneous thermal equilibrium state. The strengths $\gamma_{mn}$ of the Lindblad operator exhibits universal scaling close to a quantum phase transition for the low energy states. Furthermore, in stark contrast to STA in unitary dynamics, where control becomes most relevant close to criticality, here control is essential even away from a phase transition, as signified by non-zero values of $\gamma_{mn}$ for large energy gaps (see Eq. \eqref{eqgammn}).  This can be attributed to change in entropy $-{\rm Tr}\left(\rho(t) \ln \rho(t) \right)$ of the instantaneous thermal equilibrium state (Cf. Eq. \eqref{eq:target}) even away from criticality for a time-dependent Hamiltonian.

We have then focused on the specific example of a transverse Ising spin chain coupled to a dissipative Fermionic bath. We have considered control such that each momentum mode of the system always remains in its instantaneous Gibbs state $\rho_{G,k}^{(\beta)}$ Eq. \eqref{rhoth}. The control Lindblad operators assume a simple form in the momentum representation (Cf. Eqs. \eqref{lmn}  and \eqref{levelsk}) that may be amenable to an experimental realization. However, these operators involve multi-body interaction terms in the spin space representation, even away from criticality (see Eq. \eqref{eqT1}). We have also studied the heat current and power associated with the non-unitary dynamics. Interestingly, $\mathcal{I}_{CD} = \mathcal{P}_{CD} = 0$, signifying application of counterdiabatic Hamiltonian does not incur any operational cost. In contrast the vanishing of the energy gap close to criticality is reflected by $|\mathcal{I}_{STA}| = |\mathcal{I}_0|$ assuming maximum close to $t/\tau = \pm 1$. On the other hand, the heat current $\mathcal{I}_{STA}$  approaches zero for $|t|/\tau \gg 1$, when the steady state fails to change appreciably with time (see Fig. \ref{fig:Total_power_current}). The associated power $\mathcal{P}_{STA}$ saturates to constant value for away from criticality($|t|/\tau \gg 1$) when the energy eigenvalues $\epsilon_k$ changes with time at a constant rate $1/ \tau$.  We emphasize that even though here we have explicitly derived the results for the transverse Ising model only, however, one can follow the procedure detailed above to derive equivalent results for other free-Fermionic models as well, by considering appropriate forms of  $\Vec{b_k(\zeta(t))}$ (see Sec. \ref{secFF}).

We note that other methods of controlling excitations in systems driven through quantum critical points have also been proposed in the literature \cite{sengupta14suppressing, revathy22bath}. For example, in Ref. \cite{sengupta14suppressing} the authors showed that driving a closed quantum system across a quantum critical point using two rates, one for controlling the distance from the critical point, and another for controlling the dispersion relation of the low energy quasi-particles close to criticality, can be highly beneficial for suppressing excitations in the system. Notably, this kind of control does not require multi-particle interaction terms, as is the case for conventional STA methods in closed quantum systems \cite{delcampo12}.  However, dynamics in the presence of dissipation demands controlling the entropy of the system as well, in addition to suppressing non-adiabatic excitations due to passage through criticality. Consequently, it would be interesting to study the form of Lindblad operators required for such non-STA  control protocols \cite{sengupta14suppressing}  as well.

The control scheme presented here can be expected to be highly beneficial for the development of various quantum technologies, such as finite-time quantum heat engines involving isothermal strokes \cite{myers22quantum}; application of STA  may significantly enhance the output of such heat engines \cite{campo14more, hartmann19many, Mukherjee21}.  Furthermore, as shown in Sec. \ref{subSTAreal}, STA in real space demands multi-body interacting terms, which can be non-trivial to implement experimentally. Consequently, development of approximate STA techniques which may require interactions between fewer particles \cite{Sels17, KOLODRUBETZ20171}, can be highly relevant for experimental realization of STA in quantum critical systems in the presence of dissipation. In that respect, several currently existing platforms, such as quantum simulators \cite{sweke2016digital, muller2012engineered, georgescu2014quantum} based on  trapped ions \cite{barreiro2011open, cui2016experimental} and  Rydberg atoms \cite{kim18detailed, ebadi20quantum},  may be ideal for experimental realization of the same. In addition, our analysis suggests control in the momentum space can involve simpler terms, as compared to that in the real space (see Sec. \ref{subSTAreal}). Consequently experimental setups which allow for control in the momentum space, such as trapped-ion setups \cite{cui2016experimental}, might be suitable for experimental realization of the STA protocol studied here.

\acknowledgements
 V.M. acknowledges Adolfo del Campo for helpful discussions during various stages of the work. V.M. also acknowledges support from the University of Luxembourg, where part of the work was done, support from Science and Engineering Research Board (SERB) through MATRICS (Project No.
MTR/2021/000055) and a Seed Grant from IISER Berhampur.

\appendix

\section{General scalings in the limit $\beta \gg L^z$}
\label{appA}

For transitions from an energy level $n$ to an energy level $m \neq 1$, we have 
\ba
\gamma_{mn} &=& \frac{\dot{\lambda}_{m}(t)}{r \lambda_n(t)},\non\\
\lambda_{m} &=& \frac{e^{-\beta E_m}}{\sum_l e^{-\beta E_l}}, \non\\
\dot{\lambda}_{m} &=& \frac{-\beta \dot{E}_m e^{-\beta E_m}}{\sum_l e^{-\beta E_l}} + \frac{+\beta e^{-\beta E_m} \sum_l \dot{E}_l e^{-\beta E_l}}{\left(\sum_l e^{-\beta E_l} \right)^2} \non\\&=& \frac{-\beta \dot{E}_m e^{-\beta E_m}\left(\sum_l e^{-\beta E_l} \right) + \beta e^{-\beta E_m} \sum_l \dot{E}_l e^{-\beta E_l}}{\left(\sum_l e^{-\beta E_l} \right)^2} \non\\ 
&=& \frac{\beta e^{-\beta E_m} \sum_l e^{-\beta E_l} \left(\dot{E}_l - \dot{E}_m\right)}{\left(\sum_l e^{-\beta E_l} \right)^2}.
\ea
Again, $\lambda_n = \frac{e^{-\beta E_n}}{\sum_l e^{-\beta E_l}}$. As a result, 
\ba 
\gamma_{mn} = \frac{\beta e^{-\beta \left(E_m - E_n\right)} \sum_l^{(m)} e^{-\beta E_l} \left(\dot{E}_l - \dot{E}_m \right)}{\sum_l e^{-\beta E_l}}. 
\ea

For the lower energy states $m = 1,~n=2$ in the low temperature limit we get
\ba
\gamma_{12} &=& \frac{\beta e^{-\beta \left(E_1 - E_2\right)} \sum_l^{(1)} e^{-\beta E_l} \left(\dot{E}_l - \dot{E}_1 \right)}{\sum_l e^{-\beta E_l}} \non\\ &\approx&  \frac{\beta e^{-\beta \left(E_1 - E_2\right)} e^{-\beta E_2} \left(\dot{E}_2 - \dot{E}_1 \right)}{e^{-\beta E_1} + e^{-\beta E_2}} \non\\
&=& \frac{\beta  \left(\dot{E}_2 - \dot{E}_1 \right)}{1 + e^{-\beta \left(E_2 - E_1\right)}}, 
\ea
where we have considered $\exp\left[-\beta E_l\right] \to 0$ for $l \geq 3$.

\section{Lindblad operators and dissipation rates}
\label{appB}

The $L_{mn}(t)$ operators are as follows:
\ba
&&L_{12}^{(k)}(t) = \ket{1_t}\bra{2_t} \non\\ 
&=& \frac{\alpha_t \ket{1_k, 1_{-k}}\bra{1_k, 0_{-k}} + \ket{0_k, 0_{-k}}\bra{1_k, 0_{-k}}}{\sqrt{\left|\alpha_t^{(k)}\right|^2 + 1}}.
\label{eq:L12appB}
\ea 

We have
\ba
&&\ket{1_k, 1_{-k}}\bra{1_k, 0_{-k}} = c_k^{\dagger}c_{-k}^{\dagger}\ket{0_k, 0_{-k}}\bra{0_k, 0_{-k}}c_k \non\\
&=& c_k^{\dagger}c_{-k}^{\dagger}\Big[\ket{0_k, 0_{-k}}\bra{0_k, 0_{-k}} + \ket{1_k, 1_{-k}}\bra{1_k, 1_{-k}} \non\\ 
&+& \ket{0_k, 1_{-k}}\bra{0_k, 1_{-k}}  + \ket{1_k, 0_{-k}}\bra{1_k, 0_{-k}}        \Big]c_k\non\\
&=& c_k^{\dagger}c_{-k}^{\dagger} \mathcal{I}_k c_k = c_k^{\dagger}c_{-k}^{\dagger} c_k,
\label{eqL12_1appB}
\ea
where $ \mathcal{I}_k = \Big(\ket{0_k, 0_{-k}}\bra{0_k, 0_{-k}} + \ket{1_k, 1_{-k}}\bra{1_k, 1_{-k}} + \ket{0_k, 1_{-k}}\bra{0_k, 1_{-k}}  + \ket{1_k, 0_{-k}}\bra{1_k, 0_{-k}}\Big)$ is the identity operator corresponding to the $k$th mode. 

Similarly, 
\ba
&&\ket{0_k, 0_{-k}}\bra{1_k, 0_{-k}} = c_k c_{-k}\ket{1_k, 1_{-k}}\bra{1_k, 1_{-k}}c_{-k}^{\dagger} \non\\
&=&  c_k c_{-k}\Big[\ket{0_k, 0_{-k}}\bra{0_k, 0_{-k}} + \ket{1_k, 1_{-k}}\bra{1_k, 1_{-k}} \non\\ 
&+& \ket{0_k, 1_{-k}}\bra{0_k, 1_{-k}}  + \ket{1_k, 0_{-k}}\bra{1_k, 0_{-k}}\Big]c_{-k}^{\dagger} \non\\
&=& c_k c_{-k}\mathcal{I}_k c_{-k}^{\dagger} = c_k c_{-k}  c_{-k}^{\dagger}.
\label{eqL12_2appB}
\ea

It follows that
\ba
L_{12}^{(k)}(t) = \frac{\alpha^{(k)}_t c_k^{\dagger}c_{-k}^{\dagger} c_k + c_k c_{-k}  c_{-k}^{\dagger}}{\sqrt{|\alpha_t^{(k)}|^2 + 1}}.
\label{eqL12_3appB}
\ea 

Proceeding as above, one can show that the different Lindblad operators introduced in Eqs. \eqref{lmn} and \eqref{levelsk} are given by:
\begin{widetext}
\ba
L_{12}^{(k)}(t) &=& \left(L_{21}^{(k)}\right)^{\dagger} = \ket{1_t^{(k)}}\bra{2_t^{(k)}} = \frac{\alpha_t \ket{1_k, 1_{-k}}\bra{1_k, 0_{-k}} + \ket{0_k, 0_{-k}}\bra{1_k, 0_{-k}}}{\sqrt{\left|\alpha_t^{(k)}\right|^2 + 1}}
= \frac{\alpha^{(k)}_t c_k^{\dagger}c_{-k}^{\dagger} c_k + c_k c_{-k}  c_{-k}^{\dagger}}{\sqrt{\left|\alpha_t^{(k)}\right|^2 + 1}},\non\\
L_{13}^{(k)} &=& \left(L_{31}^{(k)}\right)^{\dagger} = \ket{1_t^{(k)}}\bra{3_t^{(k)}} = \frac{\alpha_t \ket{1_k,1_{-k}} + \ket{0_k,0_{-k}}}{\sqrt{\alpha_t^2 + 1}} \bra{0_k,1_{-k}} = \frac{\alpha_t c_k^{\dagger} c_{-k}^{\dagger} c_{-k} + c_k c_{-k} c_k^{\dagger}}{\sqrt{\alpha_t^2 + 1}},\non\\
L_{14}^{(k)} &=& \left(L_{41}^{(k)}\right)^{\dagger} = \ket{1_t^{(k)}}\bra{4_t^{(k)}} = \left(\frac{\alpha_t \ket{1_k,1_{-k}} + \ket{0_k,0_{-k}}}{\sqrt{\alpha_t^2 + 1}}\right)\left( \frac{-\theta_t \bra{1_k,1_{-k}} + \bra{0_k,0_{-k}}}{\sqrt{\theta_t^2 + 1}}\right) \non\\ &=& \frac{-\alpha_t \theta_t c_k^{\dagger} c_{-k}^{\dagger} c_k c_{-k} + \alpha_t c_k^{\dagger} c_{-k}^{\dagger} - \theta_t c_k c_{-k}  +  c_k c_{-k} c_k^{\dagger} c_{-k}^{\dagger}}{\sqrt{\left(\alpha_t^2 + 1\right) \left(\theta_t^2 + 1 \right)}}, \non\\
L_{23}^{(k)} &=& \left(L_{32}^{(k)}\right)^{\dagger} = \ket{2_t^{(k)}}\bra{3_t^{(k)}}  = \ket{1_k, 0_{-k}}\bra{0_{k},1_{-k}} = c_k^{\dagger} c_{-k}, \non\\
L_{24}^{(k)} &=& \left(L_{42}^{(k)}\right)^{\dagger} = \ket{2_t^{(k)}}\bra{4_t^{(k)}} = \ket{1_k, 0_{-k}}\frac{-\theta_t \bra{1_k,1_{-k}} + \bra{0_k,0_{-k}}}{\sqrt{\theta_t^2 + 1}} = \frac{-\theta_t c_k^{\dagger} c_k c_{-k} + c_{-k} c_k^{\dagger} c_{-k}^{\dagger}}{{\sqrt{\theta_t^2 + 1}}}, \non\\
L_{34}^{(k)} &=& \left(L_{43}^{(k)}\right)^{\dagger} = \ket{3_t^{(k)}}\bra{4_t^{(k)}} = \ket{0_k, 1_{-k}}\frac{-\theta_t \bra{1_k,1_{-k}} + \bra{0_k,0_{-k}}}{\sqrt{\theta_t^2 + 1}} = \frac{-\theta_t c_{-k}^{\dagger} c_k c_{-k} + c_{k} c_k^{\dagger} c_{-k}^{\dagger}}{{\sqrt{\theta_t^2 + 1}}}.
\label{Lks}
\ea
\end{widetext}

\begin{widetext}
The $\gamma_{mn}^{(k)}$s for the mode $k$ are given by 
\ba
\gamma_{12}^{(k)} &=& \gamma_{13}^{(k)}  = \frac{\beta e^{\beta \epsilon_k}  \left(t/\tau + \cos k \right)}{2\tau\left(1 + e^{\beta \epsilon_k} \right)\epsilon_k},\hspace{3cm}\quad
\gamma_{21}^{(k)} = -\frac{\beta e^{-\beta \epsilon_k } \left(t/\tau + \cos k \right)\tanh \left[\beta \epsilon_k/2 \right]}{4\tau \epsilon_k},\non\\
\gamma_{23}^{(k)} &=& \gamma_{32}^{(k)}=-\frac{\beta \left(t/\tau + \cos k \right) \tanh \left[\beta \epsilon_k / 2 \right]}{4\tau \epsilon_k},\hspace{1.5cm} \quad
\gamma_{31}^{(k)} = -\frac{\beta e^{-\beta \epsilon_k} \left(t/\tau + \cos k \right) \tanh\left[\beta \epsilon_k / 2 \right]}{4\tau \epsilon_k}, \non\\
\gamma_{34}^{(k)} &=& \gamma_{24}^{(k)} =  -\frac{\beta e^{\beta \epsilon_k}\left(t/\tau + \cos k \right) \tanh \left[\beta \epsilon_k/2 \right]}{4\tau \epsilon_k}, \hspace{0.89cm}\quad
\gamma_{43}^{(k)} = \gamma_{42}^{(k)} = -\frac{\beta \left( t/\tau + \cos k\right)}{2\tau \left(1 + e^{\beta \epsilon_k} \right)\epsilon_k},\non\\
\gamma_{14}^{(k)} &=& \frac{\beta e^{2\beta \epsilon_k} \left(t/\tau + \cos k \right)}{2 \tau \left(1 + e^{\beta \epsilon_k} \right)\epsilon_k}, \hspace{4cm}\quad
\gamma_{41}^{(k)} = -\frac{\beta e^{-\beta \epsilon_k} \left(t/\tau + \cos k \right)}{2\tau \left(1 + e^{\beta \epsilon_k} \right)\epsilon_k}.
\label{eq:gamk}
\ea
\end{widetext}


\section{Equation of motion:}
\label{appC}
\begin{figure}
    \centering
    \includegraphics[width=1\linewidth]{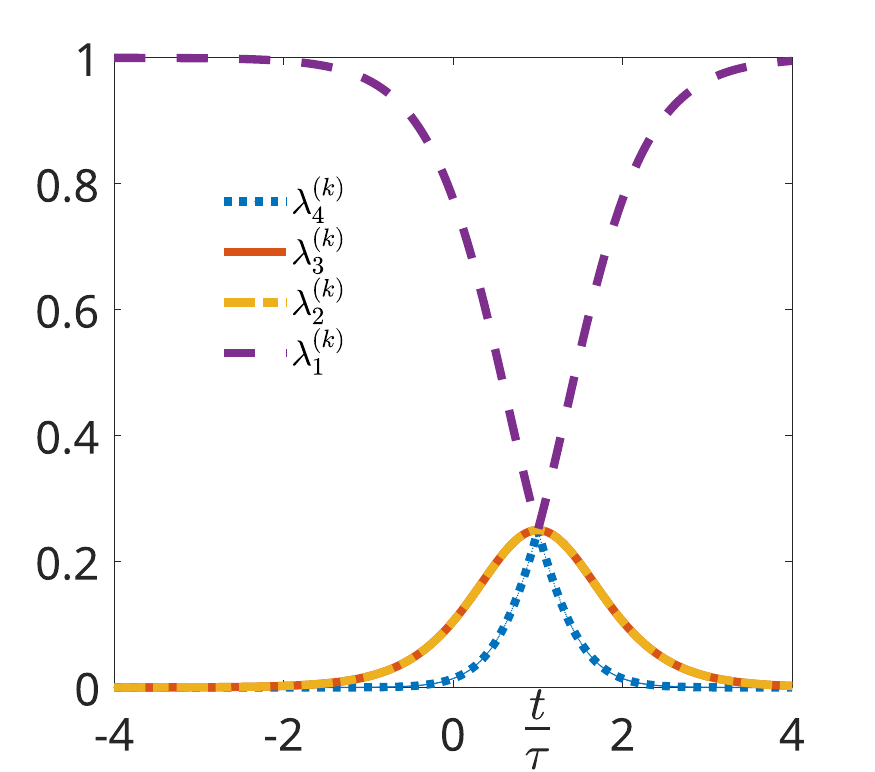}
    \caption{Thermal population $\lambda_1^{(k)}$ of the ground state, $\lambda_2^{(k)}$ and $\lambda_3^{(k)}$ of the degenerate second and third excited states, and $\lambda_4^{(k)}$ of highest excited state, with $\frac{t}{\tau}$. Here $\beta = 2 $ $k = \pi$, $\tau = 1$. The energy level populations  becomes equal at the quantum critical point $\frac{t}{\tau}=1$. }
    \label{fig:Thermal_population}
\end{figure}

The dynamical equation of motion of the system following the instantaneous thermal state (See Eq. \eqref{eq:target} and Fig. \ref{fig:Thermal_population}) can be written as \cite{Alipour20} 
\ba 
\dot{\rho}(t) &=& [H_{CD}, \rho] + \sum_{m} \dot{\lambda}_m(t) \ket{m_t} \bra{m_t}  
\label{Eq:Non-unia}\\
&=& [H_{CD}, \rho] + r \sum_{m,n}  \gamma_{mn}(t) \lambda_n (t) \ket{m_t} \bra{m_t},
\label{Eq:Non-uni}
\ea 
where we have used the relation $ \gamma_{mn} = \frac{\dot{\lambda}_m}{r \lambda_n}$. Comparing Eqs. \eqref{Eq:Non-unia} and \eqref{Eq:Non-uni} we get


\ba
\dot{\lambda}_1^{(k)}(t) &=& r \left[ \gamma_{12} \lambda_2^{(k)} (t) + \gamma_{13} \lambda_3^{(k)}(t) + \gamma_{14} \lambda_4^{(k)}(t) \right ] \non \\
\dot{\lambda}_2^{(k)}(t) &=& r \left[ \gamma_{21} \lambda_1^{(k)} (t) + \gamma_{23} \lambda_3^{(k)}(t)  + \gamma_{24} \lambda_4^{(k)} (t) \right] \non \\
\dot{\lambda}_3^{(k)} (t) &=&  r \left [ \gamma_{31} \lambda_1^{(k)} (t) + \gamma_{32} \lambda_2^{(k)}(t) + \gamma_{34} \lambda_4^{(k)}(t)  \right] \non\\
\dot{\lambda}_4^{(k)} (t)  &=& r \left[ \gamma_{41} \lambda_1^{(k)} (t) + \gamma_{42} \lambda_2^{(k)}(t) + \gamma_{43} \lambda_3^{(k)}(t) \right ]. \non
\ea 


\section{Control terms in the real space}
\label{appreal}
We focus on a particular term involved in the master equation: $\sum_k\gamma_{12}^{(k)}(L_{12}^{(k)})^{\dagger}L_{12}^{(k)} \rho$. This will lead to terms of the form 
\ba
T_1^{(12)} &=& \sum_k A^{(12)}_{1k} c_k^{\dagger} c_{-k} c_k c_k^{\dagger} c_{-k}^{\dagger} c_k \rho, 
 \\
\text{ where, }  A^{(12)}_{1k} &=& \frac{\gamma_{12}^{(k)} |\alpha^{(k)}_t|^2}{\left|\alpha_t^{(k)}\right|^2 + 1}.
\label{eq:A1kappB}
\ea
We have \cite{bunder99effect, mukherjee07quenching}
\ba
c_k = \frac{1}{\sqrt{L}} \sum_{n = 0}^L c_n e^{-ink},~~~~
c_k^{\dagger} =  \frac{1}{\sqrt{L}} \sum_{n = 0}^L c_n^{\dagger} e^{ink}.
\ea
Further, 
\ba
c_n &=& \left(\Pi_{j = 1}^{n-1} \sigma_j^z \right)(-1)^n \sigma_n^{-},\non\\
c_n^{\dagger} &=& \left(\Pi_{j = 1}^{n-1} \sigma_j^z \right)(-1)^n \sigma_n^{+}.
\ea
The $c_n, c_n^{\dagger}$ operators follow the anti-commutation relations
\ba
\{c_m, c_n^{\dagger}\} = \delta_{mn},~~~ \{c_m,c_n\} = \{c_m^{\dagger},c_n^{\dagger}\} = 0. 
\ea

Consequently,
 \begin{widetext} 
\ba
T_1^{(12)} &=&  \frac{1}{L^3} \sum_k A^{(12)}_{1k} \sum_{n,m,p,q,r,s} c_n^{\dagger} c_m c_p c_q^{\dagger} c_r^{\dagger}c_s e^{i\mathcal{D}_{1}^{(12)}k} \rho.
\ea
Assume $\{n,m,p,q,r,s\}$ such that  $n > p > r > m > q > s$. One then finds 
\ba
 \sum_k && A^{(12)}_{1k} \sum_{\{n,m,p,q,r,s\}} c_n^{\dagger} c_m c_p c_q^{\dagger} c_r^{\dagger}c_s e^{i\mathcal{D} k} \rho = -\sum_k A_{1k}^{(12)} \sum_{\{n,m,p,q,r,s\}}  c_n^{\dagger} c_p c_r^{\dagger} c_m c_q^{\dagger} c_s e^{i\mathcal{D}_1^{(12)}k} \rho \non\\
&=& \left(-1 \right)^{\mathcal{S} -1}\sum_k A_{1k}^{(12)} \sum_{n,m,p,q,r,s}  \left(\Pi_{j = 1}^{n-1} \sigma_j^z \right) \sigma_n^{+}  \left(\Pi_{j = 1}^{p-1} \sigma_j^z \right) \sigma_p^{-}  
\left(\Pi_{j = 1}^{r-1} \sigma_j^z \right) \sigma_r^{+}  \left(\Pi_{j = 1}^{m-1} \sigma_j^z \right) \sigma_m^{-} \left(\Pi_{j = 1}^{q-1} \sigma_j^z \right) \sigma_q^{+} 
\non\\&& \left(\Pi_{j = 1}^{s-1} \sigma_j^z \right) \sigma_s^{-}  e^{i\mathcal{D}_1^{(12)}k} \rho \non\\
&=&  \left(-1 \right)^{\mathcal{S} -1} \sum_k A^{(12)}_{1k} \sum_{\{n,m,p,q,r,s\}} \left(\Pi_{j = p}^{n-1} \sigma_j^z \right)   \left(\Pi_{j = m}^{r-1} \sigma_j^z \right) \left(\Pi_{j = s}^{q-1} \sigma_j^z \right)  \sigma_n^{+}  \sigma_p^{-}  
\sigma_r^{+}  \sigma_m^{-}  \sigma_q^{+} \sigma_s^{-}e^{i\mathcal{D}_{1}^{(12)}k}\rho.
\label{eqT1appB}
\ea
\end{widetext}
where $\mathcal{S} =  n + m + q + p+r+s$ and $\mathcal{D}_{1}^{(12)} = \left(n + m + q \right) - \left(p+r+s\right)$, and we have used $(\sigma_j^z)^2 = \mathcal{I}$. 

Furthermore, one can rewrite $T_1^{(12)}$ as
\begin{widetext}
    \ba
T_1^{(12)} &=&  \frac{1}{L^3} \sum_{n,m,p,q,r,s}  c_n^{\dagger} c_p c_r^{\dagger} c_m c_q^{\dagger} c_s \mathcal{J}^{(12)}_1(n,m,p,q,r,s) \rho \non\\ &=&  \frac{\left(-1 \right)^{\mathcal{S} -1}}{L^3} \sum_{\{n,m,p,q,r,s\}} \left(\Pi_{j = p}^{n-1} \sigma_j^z \right)   \left(\Pi_{j = m}^{r-1} \sigma_j^z \right) \left(\Pi_{j = s}^{q-1} \sigma_j^z \right)  \sigma_n^{+}  \sigma_p^{-}  
\sigma_r^{+}  \sigma_m^{-}  \sigma_q^{+} \sigma_s^{-} \mathcal{J}^{(12)}(n,m,p,q,r,s) \rho,\non\\
\mathcal{J}_1^{(12)}(n,m,p,q,r,s) &=& \sum_k A_{1k}^{(12)} e^{i\mathcal{D}_1^{(12)} k}.
\label{eq:JD1appB}
\ea
\end{widetext}

{\bf Term 2:} Similarly, the other non-zero term will be of the form 
\begin{widetext}
    \ba
T_2 &=& \sum_k A_{2k}  c_{-k} c_{-k}^{\dagger}c_{k}^{\dagger}  c_k c_{-k} c_{-k}^{\dagger} \rho = \frac{1}{L^3} \sum_k A_{2k}  \sum_{\{n,m,p,q,r,s\}} c_n c_m^{\dagger} c_p^{\dagger} c_q c_r c_s^{\dagger} e^{i\mathcal{D}_{2}k} 
= \frac{1}{L^3} \sum_k A_{2k} \sum_{\{n,m,p,q,r,s\}} c_m^{\dagger} c_n c_p^{\dagger} c_q c_s^{\dagger}c_r e^{i\mathcal{D}_2 k}\rho \non\\
&=& \frac{\left(-1 \right)^{\mathcal{S}}}{L^3}  \sum_k A_{2k} \sum_{\{n,m,p,q,r,s\}} \left(\Pi_{j = n}^{m-1} \sigma_j^z \right)   \left(\Pi_{j = q}^{p-1} \sigma_j^z \right) \left(\Pi_{j = r}^{s-1} \sigma_j^z \right)  \sigma_m^{+}  \sigma_n^{-}  
\sigma_p^{+}  \sigma_q^{-}  \sigma_s^{+} \sigma_r^{-}e^{i\mathcal{D}_{2}k}\rho.
\ea
\end{widetext}
Here 
\ba
A_{2k} &=& \frac{\gamma_{12}^{(k)}}{|\alpha_t^{(k)}|^2 + 1}, \non\\
\mathcal{D}_{2} &=& \left(n + p + r \right) - \left(m + q + s \right),
\ea
and we have assumed $m > n > p > q > s > r$. The remaining two terms will involve $c_k^2$ and $(c_k^{\dagger})^2$, and therefore will vanish, owing to the commutation relations satisfied by the Fermionic operators $c_k$ and $c_k^{\dagger}$.
Redefining the variables $m \to n, n \to p, p \to r, q \to m, s \to q, r \to s$, we get 
\begin{widetext}
    \ba
T_2 = \frac{\left(-1 \right)^{\mathcal{S}}}{L^3}  \sum_k A_{2k} \sum_{\{n,m,p,q,r,s\}} \left(\Pi_{j = p}^{n-1} \sigma_j^z \right)   \left(\Pi_{j = m}^{r-1} \sigma_j^z \right) \left(\Pi_{j = s}^{q-1} \sigma_j^z \right)  \sigma_n^{+}  \sigma_p^{-}  
\sigma_r^{+}  \sigma_m^{-}  \sigma_q^{+} \sigma_s^{-}e^{i\mathcal{D}_{1}^{(12)}k}\rho.
\ea
\end{widetext}

{\bf $L_{34}$ terms:} 
We note that in case of $\beta \gg L$, the most significant contributions arise from  $\gamma_{14}$ and $\gamma_{24} = \gamma_{34}$, which are responsible for deexcitation of the system to its instantaneous ground state  (see Eq. \eqref{eq:gamk2}). 

Proceeding as above, one can show that $\sum_k \gamma_{34} L_{34}^{\dagger} L_{34}$ will contain terms of the form 
\ba
T^{34}_1 &=& \sum_k A_{1k}^{34} c_{-k}^{\dagger}c_{k}^{\dagger}c_{-k}c_{-k}^{\dagger}c_k c_{-k} \rho \\&=& \frac{1}{L^3} \sum_{\{ n,m,p,q,r,s\}} c_n^{\dagger} c_m^{\dagger} c_p c_r^{\dagger} c_q c_s \sum_k A_{1k}^{(34)} e^{i \mathcal{D}_1^{(34)} k} \rho \non\\
&=& \frac{1}{ L^3} \sum_{\{ n,m,p,q,r,s\}} c_n^{\dagger} c_p c_m^{\dagger} c_q  c_r^{\dagger}  c_s \mathcal{J}_1^{(34)}(\mathcal{D}_1^{(34)}) \rho.\non
\label{eq:T34}
\ea
Here $A_{1k}^{34} = \frac{\gamma_{34} |\theta_t^{(k)}|^2}{|\theta_t^{(k)}|^2 + 1}$, $\mathcal{D}_1^{(34)} = (m + p +s) - (n + r + q)$ and in the limit of large $L$, we have
\ba
\mathcal{J}_1^{(34)}(\mathcal{D}_1^{(34)} ) = \frac{2}{\pi} \int_0^{\pi}A_{1k}^{(34)} \cos\left(\mathcal{D}_1^{(34)} k\right) dk.
\label{eq:J34}
\ea

\begin{figure}
    \centering
    \includegraphics[width=1\linewidth]{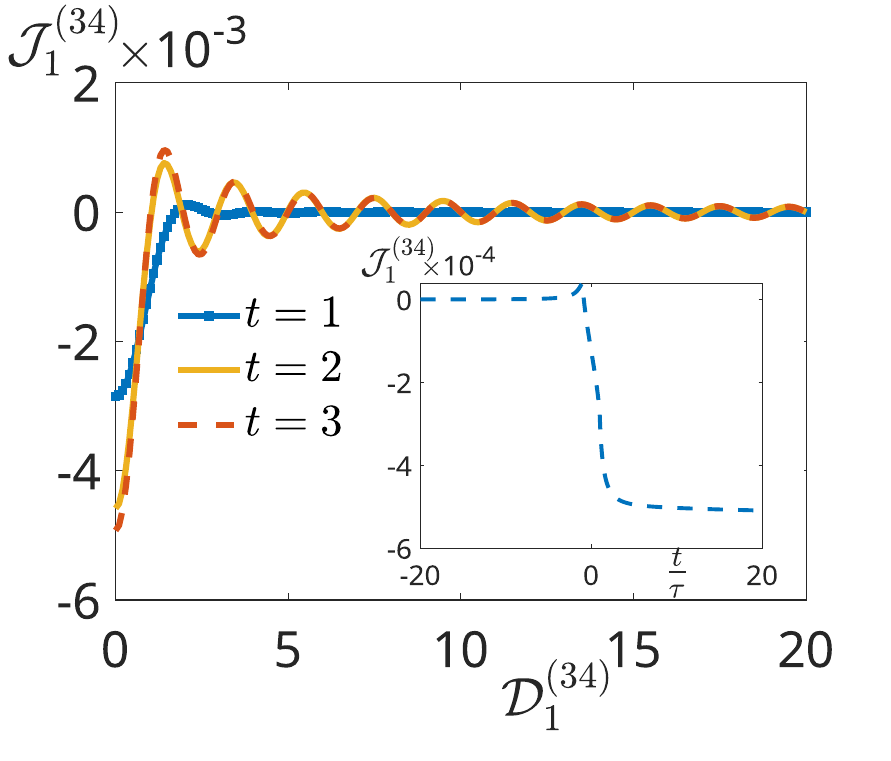}
    \caption{Variation of (a) $\mathcal{J}^{(34)}_1(\mathcal{D}_1^{(34)})$ as a function of $\mathcal{D}_1^{(34)} = \left( m + p + s \right) - \left(n+r+q\right)$ and (b) $\mathcal{J}^{(34)}_1(t)$ as a function of $\frac{t}{\tau}$ when $\mathcal{D}_1^{(34)}= 0$ . Here $\tau = 1$, $L = 1000$, $\beta = 0.01$; see Eqs. \eqref{eqT1} and \eqref{eq:J34}.}
    \label{fig:JD134}
\end{figure}

As shown in Figs. \ref{fig:JD} and \ref{fig:JD134}, both $\mathcal{J}_1^{(12)}$ and $\mathcal{J}_1^{(34)}$ decrease rapidly with $\mathcal{D}_1^{(12)}$ and $\mathcal{D}_1^{(34)}$, respectively, implying one can restrict $\mathcal{D}_1^{(12)}$ and $\mathcal{D}_1^{(34)}$ to small values only. However, we note that even for $\mathcal{D}_1^{(12)} = \mathcal{D}_1^{(34)} = 0$, we will have multi-body interaction terms in the control Lindblad operators. 
We note that by making the transformations $m \to r,~q\to m,~r\to q$ in Eq. \eqref{eq:T34} we get
\ba
T^{34}_1 &=& \frac{1}{L^3} \sum_{\{ n,m,p,q,r,s\}} c_n^{\dagger} c_p c_r^{\dagger} c_m c_q^{\dagger} c_s \non\\&& \sum_k A_{1k}^{(34)} e^{-i \mathcal{D}_1^{(12)} k}.
\ea

\section{Heat currents and Power }
\label{secheat}
\textbf{Heat Current:} We have
 \ba 
Tr([\rho, H_{STA}]H_{STA}) = Tr([\rho, H_0 &+& H_{CD}]\non\\(H_0  &+& H_{CD})).
 \ea
 Here $[\rho, H_0]=0$ by construction, while  $Tr([\rho, H_{CD}]H_{CD}) = 0$ owing to cyclic property of trace. Further,   
 \ba 
 Tr([\rho, H_{CD}]H_0) &=& Tr(\rho H_{CD}H_0 - H_{CD}\rho H_0) \non\\
 &=& Tr(\rho H_{CD}H_0) -  Tr(H_{CD}H_0\rho) \non\\&=&  Tr(\rho H_{CD}H_0) -  Tr(\rho H_{CD}H_0) \non\\ 
 &=& 0. 
 \label{I0Icd}
 \ea
Further, following Eq. \eqref{eq:heat1} we have,
\begin{widetext}
\begin{align}
\mathcal{I}_{CD} &= \text{Tr} \left[ \sum_m \dot{\lambda}_m |m_t \rangle \langle m_t| \left( \mathrm{i} \hbar \sum_n \left(|\partial_t n_t\rangle \langle n_t| - \langle n_t | \partial_t n_t \rangle |n_t \rangle \langle n_t|\right) \right) \right] \ \non \\
&= \text{Tr} \left[ \sum_m \dot{\lambda}_m |m_t \rangle \langle m_t| \left( \mathrm{i} \hbar \sum_n |\partial_t n_t \rangle \langle n_t| \right) \right] - \text{Tr} \left[ \sum_m \dot{\lambda}_m |m_t \rangle \langle m_t| \left( \mathrm{i} \hbar \langle n_t | \partial_t n_t \rangle |n_t \rangle \langle n_t| \right) \right] \ \non \\ 
&= \text{Tr} \left[ \mathrm{i} \hbar \sum_{m,n} \dot{\lambda}_m |m_t \rangle \langle m_t | \partial_t n_t \rangle \langle n_t | \right] - \text{Tr} \left[ \mathrm{i} \hbar \sum_{m,n} \dot{\lambda}_m \langle n_t | \partial_t n_t \rangle |m_t \rangle \langle n_t | ~~\delta_{mn}   \right] \ \non \\
&= \sum_{m,n} \sum_l \mathrm{i}\hbar  \dot{\lambda}_m   \langle m_t | \partial_t n_t  \rangle \langle l_t  | m_t   \rangle \langle n_t   | l_t \rangle 
- \sum_{m} \sum_l \mathrm{i}\hbar \langle l_t  | m_t \rangle \dot{\lambda}_{m} \langle m_t | \partial_t m_t  \rangle  \langle m_t | l_t \rangle  \non  \\
&= \left( \sum_m \dot{\lambda}_m  \langle \partial_t m_t | m_t  \rangle \right )  - \left(  \sum_l \dot{\lambda}_l  \langle \partial_t l_t | l_t \rangle \right) =0.
\label{eq:Icd}
\end{align}
\end{widetext}
\textbf{Power:} We evaluate the power dissipated $\mathcal{P}_{CD} = \mathrm{Tr} [ \rho(t) \dot{H}_\mathrm{CD}(t)]$ due to the counterdiabatic Hamiltonian $H_{CD}$  in presence of a non-unitary evolution by considering  the desired  path for the system to be the instantaneous thermal state  $\rho(t) = \sum_l \lambda_l(t)\ket{l_t}\bra{l_t}$ (Cf. Eq.(\ref{eq:target})).
One can use the completeness relation,  $\sum_l | l_t \rangle \langle l_t | = 1$ to get
  \begin{align}
   &\partial_t^2 \left( \sum_l | l \rangle \langle l | \right ) = \sum_l | \ddot{l} \rangle \langle l | + 2 | \dot{l} \rangle \langle \dot{l} | + | l \rangle \langle \ddot{l} | = 0 \nonumber \\
   &\Rightarrow \sum_l | \dot{l} \rangle \langle \dot{l} | =
 -\dfrac{1}{2} \sum_l \left(| l \rangle \langle \ddot{l} | + | \ddot{l} \rangle \langle l | \right).
\label{eq_trick1}
\end{align}
Further, the normalisation condition $ \langle l | l \rangle = 1$ leads to 
\begin{align}
	& \langle \dot{l} | l \rangle = - \langle l | \dot{l} \rangle ,\non
\\
      &\Rightarrow 2 \langle l | \dot{l} \rangle  = \langle l | \dot{l} \rangle - \langle \dot{l} | l \rangle.
      \label{eq_normalisation}
  \end{align}
Now we use Eq.(\ref{eq_trick1}) and Eq.(\ref{eq_normalisation}) to evaluate the time derivative of the  $H_\mathrm{CD}(t)$ (C.f Eq.(\ref{eq:Hcd})),
\begin{align}
\dot{H}_\mathrm{CD}(t)&= i \hbar \sum_l | \ddot{l} \rangle \langle l | + | \dot{l} \rangle \langle \dot{l} | - \partial_t (\langle l | \dot{l} \rangle | l \rangle \langle l |) \nonumber \\
& = \,i \hbar \sum_l \dfrac{1}{2} (\vert \ddot{l} \rangle \langle l | - | l \rangle \langle \ddot{l} |) - \partial_t (\langle l | \dot{l} \rangle) | l \rangle \langle l | \nonumber \\
&- \langle l | \dot{l} \rangle (| \dot{l} \rangle \langle l | + | l \rangle \langle \dot{l} |).
\label{eq_Hcddot}
\end{align}
Finally, we can use Eq.(\ref{eq_Hcddot}) to evaluate  $\mathcal{P}_{CD}$ as follows, 
\begin{widetext}
\begin{align}
\mathcal{P}_{CD}&= \mathrm{Tr} [ \rho(t) \dot{H}_\mathrm{CD}(t)] = i \hbar \sum_l \lambda_l \left[\dfrac{1}{2} \left(\langle l \ddot{l} \rangle - \langle \ddot{l} | l \rangle \right) - \partial_t (\langle l | \dot{l} \rangle) - \langle l | \dot{l} \rangle \left( \langle l | \dot{l} \rangle + \langle \dot{l} | l \rangle \right) \right] \nonumber \\
& = \;\dfrac{i \hbar}{2} \sum_l \lambda_l \left[ \langle l | \ddot{l} \rangle - \langle \ddot{l} | l \rangle - 2 \partial_t (\langle l | \dot{l} \rangle) \right]  
=  \; \dfrac{i \hbar}{2} \sum_l \lambda_l \left[ \langle l | \ddot{l} \rangle - \langle \ddot{l} | l \rangle - \partial_t (\langle l | \dot{l} \rangle - \langle \dot{l} | l \rangle) \right] \nonumber \\
&= \dfrac{i \hbar}{2} \sum_l \lambda_l \left[ \langle l | \ddot{l} \rangle - \langle \ddot{l} | l \rangle - \langle \dot{l} | \dot{l} \rangle - \langle l | \ddot{l} \rangle + \langle \ddot{l} | l \rangle + \langle \dot{l} | \dot{l} \rangle \right] = 0 
\end{align}
\end{widetext}

\end{document}